\documentclass[pre,aps]{revtex4}

\usepackage{ulem}

\usepackage{graphicx}
\usepackage{amsfonts}

\newlength{\figwidth}
\def\be{\begin{equation}}
\def\ee{\end{equation}}

\def\sync{synchronization }

\begin{document}

\title {\Large \bf  Detecting \sync in spatially extended discrete systems by complexity measurements}

\author{Juan R. S\'anchez$^{1}$ and Ricardo L\'opez-Ruiz$^{2}$} 
\affiliation{$^1$
Facultad de Ingenier\'{\i}a, Universidad Nacional de Mar del Plata, 
Av.J.B. Justo 4302, Mar del Plata 7600, Argentina,} 
\affiliation{$^2$
DIIS and BIFI, Facultad de Ciencias, 
Universidad de Zaragoza, 50009 - Zaragoza, Spain.}

\begin{abstract}
The \sync of two stochastically coupled one-dimensional cellular automata (CA) is analyzed. 
It is shown that the transition to \sync is characterized by a dramatic increase of 
the statistical complexity of the patterns generated by the difference automaton.  
This singular behavior is verified to be present in several CA rules 
displaying complex behavior. 
\end{abstract}

\pacs{05.45.Ra, 05.45.Xt, 07.05.Kf, 89.75.Kd} 

\maketitle

Despite all the efforts devoted to understand the meaning of {\it complexity},
we still do not have an instrument in the laboratories specially designed for
quantifying this property. Maybe this is not the final objective of all those theoretical 
attempts carried out in the most diverse fields of knowledge in the last years
\cite{grassberger,lloyd,shiner,kolmogorov,chaitin,lempel,bennett,crutchfield},
but, for a moment, let us think in that possibility.

Similarly to any other device, our hypothetical apparatus will have an input
and an output. The input could be the time evolution of some variables 
of the system. The instrument records those signals, analyzes 
them with a proper program and finally screens the result in the form of 
a {\it complexity measurement}.
This process is repeated for several values of the 
parameters controlling the dynamics of the system.
If our interest is focused in the {\it more complex configuration} of the system 
we have now the possibility of tuning such an state by regarding the complexity 
plot obtained at the end of this process.

As a real applicability of this proposal, let us apply it to an {\it  \`a-la-mode} problem.
The clusterization or \sync of chaotic coupled elements was put in evidence at the beginning 
of the nineties \cite{kaneko,lopez91}. Since then, a lot of publications have been devoted
to this subject \cite{boccaletti}. Let us consider one particular of these systems to illuminate
our proposal.

(1) SYSTEM:
We take two coupled elementary one dimensional cellular
automata displaying complex spatio-temporal dynamics \cite{wolfram}. 
Recently, it has been show that this system can undergo through a \sync transition \cite{zanette}. 
The transition to full \sync occurs at a critical value $p_c$ of a \sync
parameter $p$. Briefly the numerical experiment
is as follows. Two $L$-cell cellular automata (CA) with the same evolution rule $\Phi$ are started
from different random initial conditions for each automaton. Then, at each time
step, the dynamics of the coupled CA is governed by the successive
application of two evolution operators; the independent evolution of each CA according
to its corresponding rule $\Phi$ and the application of a stochastic operator that compares
the states $\sigma_i^1$ and $\sigma_i^2$ of all the cells, $i=1,...L$, in each
automaton. If $\sigma_i^1=\sigma_i^2$, both states are kept invariant. 
If $\sigma_i^1\neq\sigma_i^2$, they are left unchanged 
with probability $1-p$, but both states are updated either to $\sigma_i^1$
or to $\sigma_i^2$ with equal probability $p/2$.
It is shown in reference \cite{zanette} that there exists
a critical value of the \sync parameter ($p_c=0.193$ for the rule $18$) 
above which full \sync is achieved.  

(2) DEVICE: 
We choose a particular instrument to perform our measurements, 
that capable of displaying the value of the {\it LMC complexity} ($C$) \cite{LMC}.
The statistical complexity $C$ is defined as follows,
\begin{eqnarray}
C(\{\rho_i\}) & = & H(\{\rho_i\})\cdot D(\{\rho_i\})\; = \nonumber \\
    & = & -k \left[ \sum_{i=1}^N \rho_i\log \rho_i \right]\times \:
    \left[ \sum_{i=1}^N \,\left( \rho_i - \frac{1}{N} \right)^2\right ]\:,
    \label{eq:def-c}
\end{eqnarray}      
where $\{\rho_i\}$ represents the set of probabilities of the $N$ accessible 
discrete states of the system, 
with $\rho_i\geq 0$ , $i=1,\cdots,N$, and $k$ is a constant.
If $k=1/logN$ then we have the normalized complexity.
$C$ is a statistical measure of complexity that identifies the
entropy or information stored in a system and its disequilibrium,
i.e., the distance from its actual state to the 
probability distribution of equilibrium, as the two basic ingredients for calculating
the complexity of a system. 
This quantity vanishes both for completely ordered and for completely 
random systems giving then the correct asymptotic properties required for
a such well-behaved measure. 
The calculation of $C$ has been useful to successfully discern many situations 
regarded as complex in discrete systems out of
equilibrium \cite{calbet,martin,guozhang,zuguo,lovallo}. 

(3) INPUT:
In particular the evolution of two coupled CA evolving under 
the rules $22$, $30$, $90$ and $110$ is analyzed. 
The pattern of the difference automaton will be the input of our device.
In Fig. 1 it is shown it for a coupling probability $p=0.23$,
just above the synchronization transition.
The left and the right plots show 
$250$ successive states of the two automata,
whereas the central plot displays the corresponding difference automaton.
Such automaton is constructed by comparing one by one all the sites ($L=100$)
of both automata and putting zero when the states $\sigma_i^1$
and $\sigma_i^2$, $i=1\ldots L$, are equal or putting one otherwise. 
It is worth to observe that 
the difference automaton shows an interesting {\it complex
structure} close to the \sync transition. This complex pattern
is only found in this region of parameter space. 
When the system if fully synchronized the difference automaton
is composed by zeros in all the sites, while when there is no \sync at all 
the structure of the difference automaton is completely random.

(4) METHOD OF MEASUREMENT:
How to perform the measurement of $C$ for such two-dimensional patterns has been 
presented recently in Ref. \cite{LR-S}. 
We let the system evolve until the asymptotic regime is attained. 
The variable $\sigma_i^d$ in each cell of the difference pattern  
is successively translated to an unique binary sequence
when the variable $i$ covers the spatial dimension of the lattice, 
$i=1,\ldots,L$, and the time variable $n$ is consecutively increased.
This binary string is analyzed in blocks of $n_o$ bits, where
$n_o$ can be considered the scale of observation.
The accessible states to the system among the $2^{n_o}$ possible states 
is found as well as their probabilities.
Then, the magnitudes $H$, $D$ and $C$ are directly calculated
and screened by the device.   

(5) OUTPUT:
The results of the measurement are shown in Fig. 2.
The normalized complexity $C$ as a function of the \sync parameter $p$ is plotted for 
different coupled one-dimensional CA that evolve under the rules $22$, $30$, $90$ 
and $110$ , which are known to generate complex patterns. 
All the plots of Fig. 2 were obtained using the following parameters: number of 
cell of the automata, $L=1000$; total evolution time, $T=600$ steps.
For all the cases and scales analyzed, the statistical complexity $C$
shows a dramatic increase close to the \sync transition. It reflects the complex structure
of the difference automaton and the capability of the measurement device here  
proposed for {\it clearly signaling}  the \sync transition of two coupled CA. 

These results are in agreement with the measurements of $C$
performed in the patterns generated by a one-dimensional logistic coupled map
lattice in Ref. \cite{LR-S}. There the {\it LMC statistical complexity} ($C$)  
also shows a singular behavior close to the two edges 
of an absorbent region where the lattice displays spatio-temporal intermittency.
Hence, in our present case, the \sync region of the coupled systems can be interpreted
as an absorbent region of the difference system.
In fact, the highest complexity is reached on the border of this 
region for $p\approx 0.2$. The parallelism between both systems is therefore complete.
 
Finally, let us remark that the critical parameter for switching \sync in these 
systems is similar for all of them ($p_c\approx 0.2$). 
This fact could deserve some operational explanation in a future work.

\newpage

\begin{figure}[h]
  \centering
  {\includegraphics[angle=0, width=15cm]{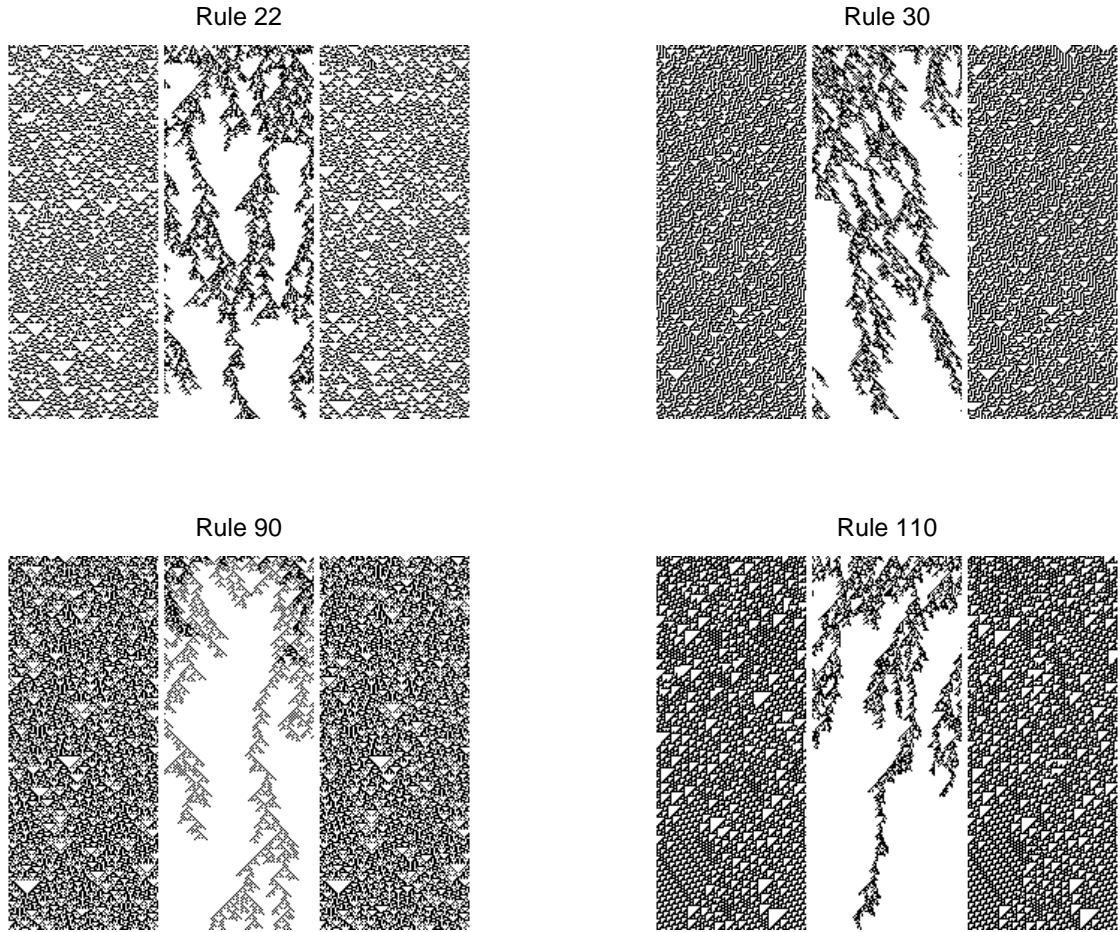}}
  \caption{Spatio-temporal patterns just above the \sync transition.
The left and the right plots show $250$ successive states of the two coupled 
automata and the central plot is the corresponding difference automaton
for the rules $22$, $30$, $90$ and $110$. The number of sites is $L=100$
and the coupling probability is $p=0.23$.}
  \label{fig1}
\end{figure}

\begin{figure}[h]
  \centering
  {\includegraphics[angle=0, width=15cm]{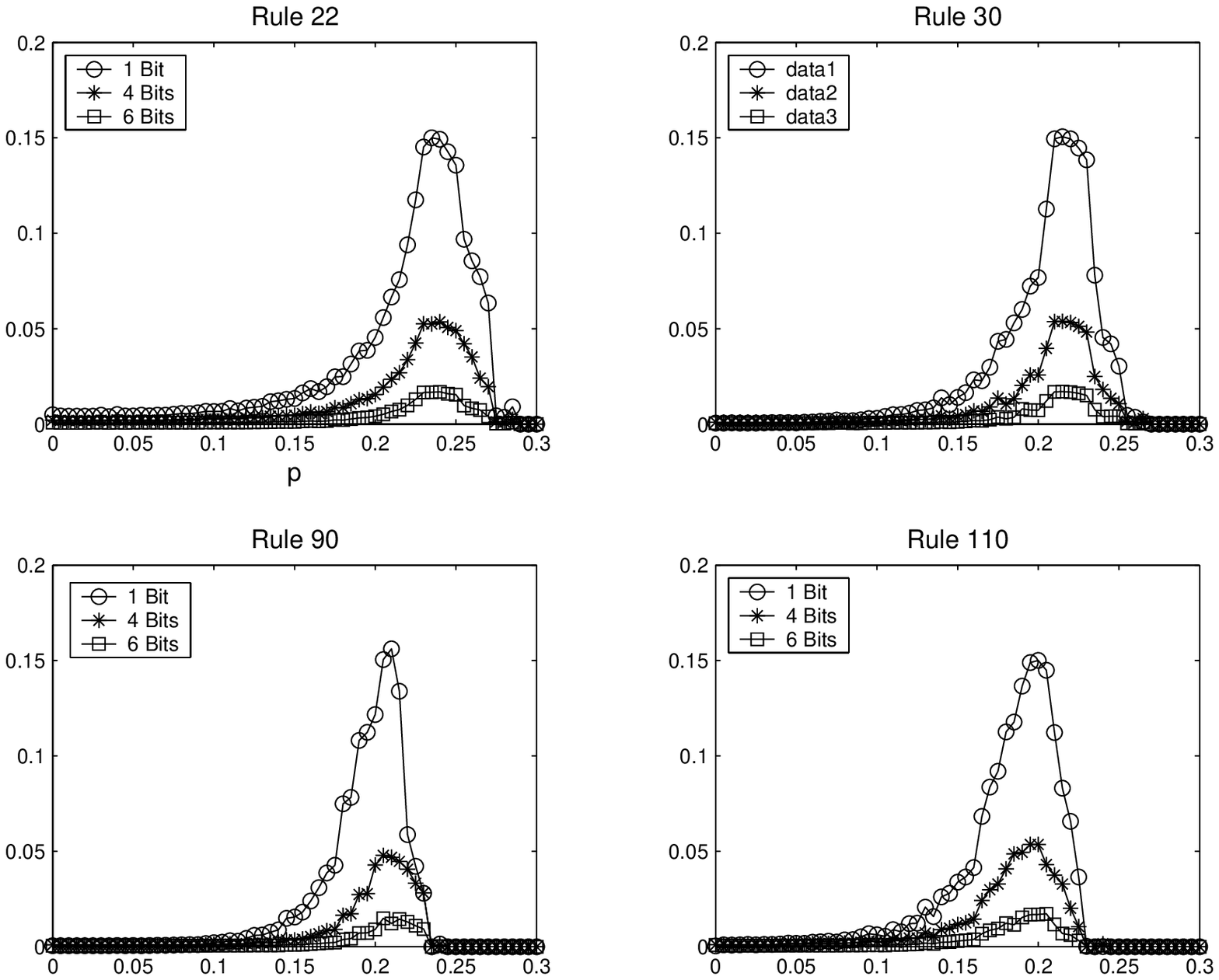}}
  \caption{The normalized complexity $C$ versus the coupling probability $p$ 
  for different scales of observation: $n_o=1$ ($\circ$),$n_o=4$ ($\star$) 
  and $n_o=6$ ($\Box$). $C$ has been calculated over the last $300$ iterations 
  of a running of $600$ of them for a lattice with $L=1000$ 
  sites. The \sync transition is clearly depicted 
  around $p\approx 0.2$  for the different rules.}
  \label{fig2}
\end{figure}


\begin{thebibliography}{99}

   \bibitem{grassberger}
    P. Grassberger, Int. J. Theor. Phys. {\bf 25}, 907 (1986).

    \bibitem{lloyd}
    S. Lloyd and H. Pagels, Ann. Phys. (NY) {\bf 188}, 186 (1988).

    \bibitem{shiner}
    J.S. Shiner, M. Davison and P.T. Landsberg, Phys. Rev. E {\bf
    59}, 1459 (1999).

    \bibitem{kolmogorov}
    A.N. Kolmogorov, Probl. Inform. Theory {\bf 1}, 3 (1965).

    \bibitem{chaitin}
    G. Chaitin, J. Assoc. Comput. Mach. {\bf 13}, 547 (1966).

    \bibitem{lempel}
    A. Lempel and J. Ziv, IEEE Trans. Inform. Theory {\bf 22},
	75 (1976).

	\bibitem{bennett}
	C.H. Bennett, in {\it Emerging syntheses in science}, 
	Ed. D. Pines, Addisson-Wesley, Reading MA (1988).
	
    \bibitem{crutchfield}
    J.P. Crutchfield and K. Young, Phys. Rev. Lett. {\bf 63}, 
	105 (1989).
	
	\bibitem{kaneko}
	K. Kaneko, Phys. Rev. Lett. {\bf 63}, 219 (1989); Physica D {\bf 41}, 137 (1990).
	
	\bibitem{lopez91}
	R. L\'opez-Ruiz and C. P\'erez-Garcia, Chaos, Solitons and Fractals {\bf 1}, 511 (1991).
	
	\bibitem{boccaletti}
	S. Boccaletti, J. Kurths, G. Osipov, D.L. Valladares and C.S. Zhou, 
	Phys. Rep. {\bf 366}, 1 (2002).

	\bibitem{wolfram}
	S. Wolfram, Rev. Mod. Phys. {\bf 55}, 601 (1983).
	
	\bibitem{zanette} 
	L. G. Morelli and D. H. Zanette, Phys. Rev. E {\bf 58}, R8 (1998).

	\bibitem{LMC} 
	R. L\'{o}pez-Ruiz, H.L. Mancini and X. Calbet, Phys. Lett. A {\bf 209}, 321 (1995).

	\bibitem{calbet} 
	X. Calbet and R. L\'{o}pez-Ruiz, Phys. Rev. E {\bf 6}, 066116(9) (2001).

	\bibitem{martin} 
	M.T. Martin, A. Plastino and O.A. Rosso, Phys. Lett. A {\bf 311} (2-3), 126 (2003). 

	\bibitem{guozhang} 
	G. Feng and S. Song and P. Li, J. Hydr. Eng. Chin. (Hydr. Eng. Soc.) {\bf 11}, article no. 14 (1998).
	
    \bibitem{zuguo}
	Z. Yu and G. Chen, Comm. Theor. Phys. (Beijing China) {\bf 33} 673 (2000).
	
	\bibitem{lovallo}
	M. Lovallo, V. Lapenna and L. Telesca, Chaos, Solitons and Fractals {\bf 24}, 33 (2005).

	\bibitem{LR-S} 
	J. R. S\'anchez and R. L\'{o}pez-Ruiz. nlin.PS/0410062, submitted for publication (2004).


\end{thebibliography}
\end{document}